\documentclass[conference]{IEEEtran}
\usepackage{subcaption}

\ifCLASSINFOpdf
  \usepackage[pdftex]{graphicx}
\else
   \usepackage[dvips]{graphicx}
\fi
\usepackage{amssymb}
\usepackage[cmex10]{amsmath}
\begin{document}
%
\title{Personalising Mobile Advertising Based on Users' Installed Apps}


\author{\IEEEauthorblockN{Jenna Reps, Uwe Aickelin and Jonathan Garibaldi}
\IEEEauthorblockA{School of Computer Science \& Advanced Data Analysis Center\\
University of Nottingham\\
Nottingham, NG8 1BB\\
Email: \{jenna.reps, uwe.aickelin, jon.garibaldi\}@nottingham.ac.uk}
\and
\IEEEauthorblockN{Chris Damski}
\IEEEauthorblockA{Opera Mediaworks \\ The Tower Building \\ London, SE1 7NX}
}


%


\maketitle

\begin{abstract}
Mobile advertising is a billion pound industry that is rapidly expanding. The success of an advert is measured based on how users interact with it. In this paper we investigate whether the application of unsupervised learning and association rule mining could be used to enable personalised targeting of mobile adverts with the aim of increasing the interaction rate.  Over May and June 2014 we recorded advert interactions such as tapping the advert or watching the whole advert video along with the set of apps a user has installed at the time of the interaction.  Based on the apps that the users have installed we applied k-means clustering to profile the users into one of ten classes.  Due to the large number of apps considered we implemented dimension reduction to reduced the app feature space by mapping the apps to their iTunes category and clustered users based on the percentage of their apps that correspond to each iTunes app category.  The clustering was externally validated by investigating differences between the way the ten profiles interact with the various adverts genres (lifestyle, finance and entertainment adverts).  In addition association rule mining was performed to find whether the time of the day that the advert is served and the number of apps a user has installed makes certain profiles more likely to interact with the advert genres.    The results showed there were clear differences in the way the profiles interact with the different advert genres and the results of this paper suggest that mobile advert targeting would improve the frequency that users interact with an advert.

\end{abstract}


%
\IEEEpeerreviewmaketitle

\section{Introduction}

Mobile advertising is a rapidly expanding  industry. In 2013 global mobile ad spend was estimated at \$17.96 billion, rising to a predicted \$31.45 billion in 2014 \cite{Emarketer2014}. In the UK, the mobile advertising industry is worth \textsterling 1.031 billion - accounting for 16.3\% of all UK digital ad spending \cite{Iab2014}. This growth is partially fuelled by smartphone adoption, with the International Data Corporation (IDC) measuring overall smartphone market growth at 23.1 per cent year on year \cite{Idc2014}.

Opera Mediaworks operates the world's biggest brand-focused mobile ad network, providing 24 of the 25 top global brands with a complete mobile advertising package. This means it operates both on the supply side, by representing publishers, and the demand side, by working with advertisers to deliver targeted media solutions. Opera Mediaworks represents over 14,000 publishers globally with over half a billion unique monthly customers serving over 64 billion ad impressions per month.

It is in the interests of both the ad publisher and the brand to receive a high interaction rate, and as such personalised ad targeting is vital as it increases the likelihood that a person viewing the ad clicks through. Ad targeting identifies associations between personal attributes such as age, gender or occupation and the probability of a user interacting with an ad impression (when an advert is displayed to a user it is referred to as an impression). It is also possible to group people with similar attributes into clusters and identify the kinds of apps they download on their mobile. The apps that a user has installed can then be used to determine how they would interact with mobile ads.

In previous work researchers have investigated using personal information extracted from social media sites such as Facebook to enable personalised mobile ad targeting \cite{shannon2009profiling} or using personal information stored on a phone such as email and web browsing history \cite{haddadi2010mobiad}.  To the best of our knowledge, there has been no research to date investigating whether people can be successfully targeted for an ad based on the apps they have installed onto their mobile device.  Therefore, in this paper we use clustering techniques to cluster people into one of ten profiles based on their installed apps and then investigate the differences between interacting with finance, entertainment and lifestyle ads for the various profiles.  In addition, we found associations between user profile, the time of the day of advert interaction, the total number of apps downloaded and the type of ad interaction. These associations were used to demonstrate how different types of users could be targeted depending on the time of day.

This continuation of this paper is as follows.  In section \ref{mat} we described the data collection and preprocessing followed by the details of the analysis performed. The results of the analysis are presented and discussed in section \ref{res} and the paper concludes with section \ref{conc}.

\section{Materials \& Methods} \label{mat}
\subsection{Data Collection}
To analyse associations between a user's installed app profile and his interaction across various ad categories the ad interactions of millions of users were logged over a study period.   The process of interacting with an ad depends on the type of ad.  For standard ads, the user must click on the ad after being served the ad impression and then the user can play the ad's video until they either finish watching the video or tap on the video exit button.  Some ads are auto playing, for these ads the user will interact by continuing to view the video or taping the video exit. In general, the recorded interaction stages are: impression, load video, play video, video 25\%, video 50\%, video 75\% and complete video .  However, the user can exit, pause and restart the video at any point during its progression. 

For each ad interaction (e.g. impression, play video, complete video) the log file contained:
\begin{itemize}
\itemsep0em 
\item Unique user ID - this is an anonymous user ID used to identify when users have multiple impressions for the same ad.
\item App list - this is the set of apps that a user has installed on their mobile device at the time of the ad interaction
\item Interaction date -the date and and time of the ad interaction 
\item Advert- the name of the advertisement campaign that the user interacted with 
\item Type of interaction (impression, play video, watch 50\% of video, etc.)
\item Publisher - the name of the publisher
\item Publisher site - the name of the publisher's site where the ad is being displayed to the user
\end{itemize}

The data were collected from the 2nd of May 2014 until 22nd of June 2014.  In total there were 60,696 log files consisting of approximately 10 million ad interactions for over 2 million people.  During this time there was ten different types of ads (i.e., ten advertisement campaigns).  The ads were classed as either an entertainment ad, a lifestyle ad or a finance ad.  For this study we only considered users with an iPhone device.

\begin{table*} \centering
\begin{tabular}{ccccccc}
user ID & App list & Date & Interaction & Advert & Publisher & Site \\ \hline
U7 & entertainment1,finance5,finance67,lifestyle78 & 03/05/2014 8:00pm & Impression & Advert1 & Pub6 & Site2 \\
U7 & entertainment1,finance5,finance67,lifestyle78 & 05/05/2014 4:03pm & Impression & Advert4 & Pub2 & Site1 \\
U7 &entertainment1,finance5,finance67,lifestyle78 & 05/05/2014 4:03pm & Tap & Advert4 & Pub2 & Site1 \\
U7 & entertainment1,finance5,finance67,lifestyle78 & 05/05/2014 4:04pm & Load Video & Advert4 & Pub2 & Site1 \\
U7 & entertainment1,finance5,finance67,lifestyle78 & 05/05/2014 4:04pm & Play Video & Advert4 & Pub2 & Site1 \\
U7 & entertainment1,finance5,finance67,lifestyle78 & 05/05/2014 4:05pm & 25\% Video & Advert4 & Pub2 & Site1 \\
U7 & entertainment1,finance5,finance67& 15/06/2014 6:75pm & Impression & Advert1 & Pub6 & Site2 \\
U23 & finance1, entertainment34, entertainment33, finance4, lifestyles3, entertainment6 & 21/06/2014 2:18am & Impression & Advert4 & Pub4 & Site1 \\
\end{tabular}
\caption{A made up example of the data collected.}
\label{exp}
\end{table*}

Table \ref{exp} illustrates an example of the data collected.  Additional information about the site and publisher where the ad was served was also recorded in case these caused any bias. For example, the ad's video download speed may very depending on the publishers serving the ad and this may influence whether users play the video.  

A user can be served the same ad impression multiple times and this can cause issues when looking for associations between the installed apps and ad interactions as a user is unlikely to play a video that he or she already watched. To reduce this bias we only considered the first occurrence of the interaction (e.g., first impression, first time the video was loaded, first time the video was played). 

\subsection{Analysis}
\subsubsection{Profiling Users}
To profile the users we preprocessed the app data to reduce the number of features and then implemented k-means clustering with k=10.  This number was chosen as preliminary work investigating the internal validation measures such as within-cluster sum of square distance and between-cluster sum of square distance indicated that the clustering performance increased as the number of clusters increased.  However, with k$>10$, many of the clusters were very small.  This is undesirable from a marketing perspective, as it means targeting only a small number of the population. There were a total of 820 apps that were included in the study.  For the $i$th user and $j$th app, the value
\begin{equation*}
x_{ij} = \left\{ 
  \begin{array}{l l}
    1 & \quad \text{if app $j$ was installed anytime during the study}\\
    0 & \quad \text{otherwise}
  \end{array} \right.
\end{equation*}
represents whether the user had the $j$th app installed at any point during the study.  The vector $\mathbf{x_{i}} \in \{0,1\}^{820}=(x_{i1}, x_{i2}, ..., x_{i820})$ is a binary vector representing the apps that user $i$ had installed during the study.

Clustering with 820 binary features (the vectors $\mathbf{x_{i}}$) is likely to cause issues due to the curse of dimensionality \cite{bellman1961adaptive} and it would also be computationally very expensive.  To overcome this issue we reduce the number of features by using domain knowledge.  To reduce the feature space we considered the app genres rather than the actual apps.  For each user, we mapped the users apps into their corresponding iTunes categories and then calculated the number of apps of each category that the user has installed.  We then normalised this by dividing by the user's total number of apps.  This effectively gives us the percentage of the user's total apps that correspond to each category.  As there are 22 categories, this reduced the feature space from 820 to 22.  

For the 22 iTune's app categories, we define the category vectors as,
\begin{equation*}
C_{kj} = \left\{ 
  \begin{array}{l l}
    1 & \quad \text{if app $j$ belongs to category $k$}\\
    0 & \quad \text{otherwise}
  \end{array} \right.
\end{equation*}
where the vector $\mathbf{C_{k}} \in \{0,1\}^{820}=(C_{k1}, C_{k2}, ..., C_{k820})$ is a binary vector representing which of the 820 apps belong to the iTune's $k$th category.  The mapping from the apps vector to the iTunes category vector is then the vector of the dot product for each category vector with the user's app vector divided by the total number of apps installed by the user,
\begin{equation*}
\hat{x}_{i} = (\mathbf{x_{i}}.\mathbf{C_{1}}, \mathbf{x_{i}}.\mathbf{C_{2}}, ..., \mathbf{x_{i}}.\mathbf{C_{22}} )/N_{j}
\end{equation*}
where $\mathbf{x_{i}}.\mathbf{C_{k}}= \sum_{j} x_{ij}C_{kj}$ and $N_{j}=\sum_{j} x_{ij}$.

For example, consider the user U23 in Table \ref{exp}, their app list is (finance1, entertainment34, entertainment33, finance4, lifestyles3, entertainment6), so their category percentage would be finance: 2/6, entertainment: 3/6, lifestyles: 1/6.  As users change their apps over time, but their uninstalled apps are still insightful,  we considered a user's app lists to be all the apps they had during the study, rather than just the ones they had at the time of the interaction.

Once the app feature space had been mapped into the category percentage feature space, $ f(\mathbf{x_{i}}): \{0,1\}^{820} \to \mathbb{R}^{22}; f(\mathbf{x_{i}})=\hat{\mathbf{x_{i}}}$, we then normalised the data by calculating the standard score (subtract the mean and divide by the standard deviation) to ensure each iTune's category is equally treated \cite{milligan1988study} and implemented k-means clustering \cite{hartigan1979algorithm} on the set $X=\{\hat{\mathbf{x}_{i}} \}$ with k=10.  The internal validity of the clusters were investigated by calculating the within-cluster and between-cluster sum of squares, the sum of the squared difference between each data-point in the same cluster and the sum of the squared difference between each data-point in different clusters, respectively.   

The differences between the ways the ten cluster profiles interact with each ad genre were then investigated.  The index value of each cluster profile's interaction is calculated as the frequency of the interaction within the cluster profile divided by the frequency of the interaction for all the users.  Using the results of the k-means clustering, for $K \in \mathbb{N}_{\le10}$ we denote
\begin{equation*}
y^{K}_{i} = \left\{ 
  \begin{array}{l l}
    1 & \quad \text{if user $i$ belongs to cluster $K$}\\
    0 & \quad \text{otherwise}
  \end{array} \right.
\end{equation*}
where $\mathbf{y}^{K}=(y^{K}_{1}, y^{K}_{2}, ..., y^{K}_{n})$ is a binary vector specifying which of the users are in the $K$th cluster.  For example, $\mathbf{y}^{1}=(y^{1}_{1}=1, y^{1}_{2}=1, ..., y^{1}_{n}=0)$ means that the first and second users are in the first cluster, whereas the nth user is in a different cluster. The total number of users within cluster $K$ is denoted by $|\mathbf{y}^{K}|$.

The vector detailing the users who received an impression for the $L$th advert is,
\begin{equation*}
A^{L0}_{i} = \left\{ 
  \begin{array}{l l}
    1 & \quad \text{if user $i$ had an impression of the $L$th advert}\\
    0 & \quad \text{otherwise}
  \end{array} \right.
\end{equation*} 
where $\mathbf{A}^{L0}=(A^{L0}_{1}, A^{L0}_{2}, ..., A^{L0}_{n})$ is a binary vector specifying which of the users had an impression of the $L$th advert. Similarly, the vector detailing the users who interacted with (e.g. played the advert's video) the $L$th advert is,
\begin{equation*}
A^{L}_{i} = \left\{ 
  \begin{array}{l l}
    1 & \quad \text{if user $i$ interacted with the $L$th advert}\\
    0 & \quad \text{otherwise}
  \end{array} \right.
\end{equation*} 
where $\mathbf{A}^{L}=(A^{L}_{1}, A^{L}_{2}, ..., A^{L}_{n})$ is a binary vector specifying which of the users interacted with the $L$th advert.  The index value of cluster $K=j$ interacting with a set of adverts $S$ is, 
\begin{equation}
Ind_{K}= \frac{\sum_{L \in S}(\mathbf{y}^{K}.\mathbf{A}^{L})/  \sum_{L \in S}(\mathbf{y}^{K}.\mathbf{A}^{L0})}{\sum_{L \in S, k\in \mathbb{N}_{\le10}}(\mathbf{y}^{k}.\mathbf{A}^{L})/  \sum_{L \in S, k\in \mathbb{N}_{\le10}}(\mathbf{y}^{k}.\mathbf{A}^{L0})}
\end{equation}

\subsubsection{Association Rule Mining}
 In addition to investigating the index value we also investigate associations between the users' profiles, the time of the day of the interaction and the users' total number of apps.  

Association rule mining finds relationships of the form `antecedent' $\to$ `consequence', where the `antecedent' is associated to the `consequence'.  The most commonly implemented algorithm for association rule mining is the Apriori algorithm \cite{agrawal1996fast}. This algorithm works by restricting the search space to itemsets that occur frequently, satisfying some minimum support constraint, and finding rules containing these itemsets with a sufficiently high confidence (an approximation of the conditional probability of the `consequence' occurring given the `antecedent' occurs).

Formally, consider a database as a set of baskets, $D=\{b_{1}, b_{2}, .., b_{n}  \}$, where each basket is a subset of $I$, a set of items $I=\{i_{1}, i_{2}, ..., i_{m} \}$. The support of itemset $A$, denoted Supp($A$), is the number of baskets in the database that contain the itemset, $|X \in D  : A \subset X|$. The support of rule $A \to B$ is the number of baskets that contain both the antecedent and the consequence, $|X \in D  : A \cup B \subset X|$.  When the consequence is rare, the minimum support confidence needs to be set extremely low to get rules of interest and this often causes algorithm efficiently issues.  To overcome this problem, the left support can be used as a constraint instead.  The left support is simply the support of the antecedent. The confidence of rule $A \to B$, denoted Conf($A \to B$), is the number of baskets that contain both $A$ and $B$ divided by the number of baskets that contain $A$, $Supp(A \cup B)/Supp(A)$. The lift is a measure of dependancy, $Lift(A \to B)= Supp(A \cup B)/[Supp(A)Supp(B)]$.  This gives a value of how much more likely the consequence is to occur after the antecedent than in general and is similar to the index value measure. 

Previously the users were clustered using k-means into one of ten cluster profiles.  We also classified the users into four groups based on how many apps they had installed on their device over the study.  Denoting $\bar{x}$ as the mean number of apps that users have and $\sigma$ as the standard deviation of the number of apps that users have, the total installed app classification was,
\begin{description}
\item[class1]: total number of apps $ \in [0 , \bar{x}-2\sigma)$
\item[class2]: total number of apps $ \in [\bar{x}-2\sigma, \bar{x}-\sigma)$
\item[class3]: total number of apps $ \in [\bar{x}-\sigma, \bar{x}+\sigma)$
\item[class4]: total number of apps $ \in [\bar{x}+2\sigma, \infty)$
\end{description} 
and we classified the time of the day that the impression occurred as,
\begin{description}
\item[night]: time  $ \in [00:00 , 06:00) \cup [22:00,00:00]$ 
\item[daytime]: time  $ \in [06:00, 17:00)$
\item[evening]: time $ \in [17:00, 22:00)$
\end{description}

To find the association rules we created `baskets' containing the user's cluster profile, the user's total installed apps class, the user's time of day class for the impression and the set of interactions that followed the impression.  For example, using Table \ref{exp}, an example basket for user U7 for advert4 is \{cluster5,  class1, daytime, impression, tap, loadvideo, playvideo, video25 \} assuming U7 was found to be in cluster profile 5 and he only had 4 apps recorded during the study which is very low.  

For a specific ad or ad genre we found all the users served an impression and created their baskets as detailed above.  We then implemented association rule mining with a minimum left support set at 1 $\times 10^{-5}$, a minimum confidence equal to the support of the consequent item of interest and filtered the rules with a lift greater than 1.5 (this means the users with the rule's antecedent were 50\% more likely to have the consequence than in general).  We put a constraint on the rules to only mine those with the consequence of either `playing the video' or `watching up the 50\%' of the video or `watching 100\%' of the video.

\subsubsection{Software}
The data were stored and access using SQL and the analysis was performed using the freely available open source software R \cite{r}.  The R libraries used for the association rule mining was arules \cite{arules}, the library stats \cite{r} was used for the k-means clustering.

\section{Results \& Discussion} \label{res}
\subsection{Profiles}
\begin{figure*}
\centering
\includegraphics[trim = 0mm 0mm 0mm 0mm, clip,width=0.9\textwidth]{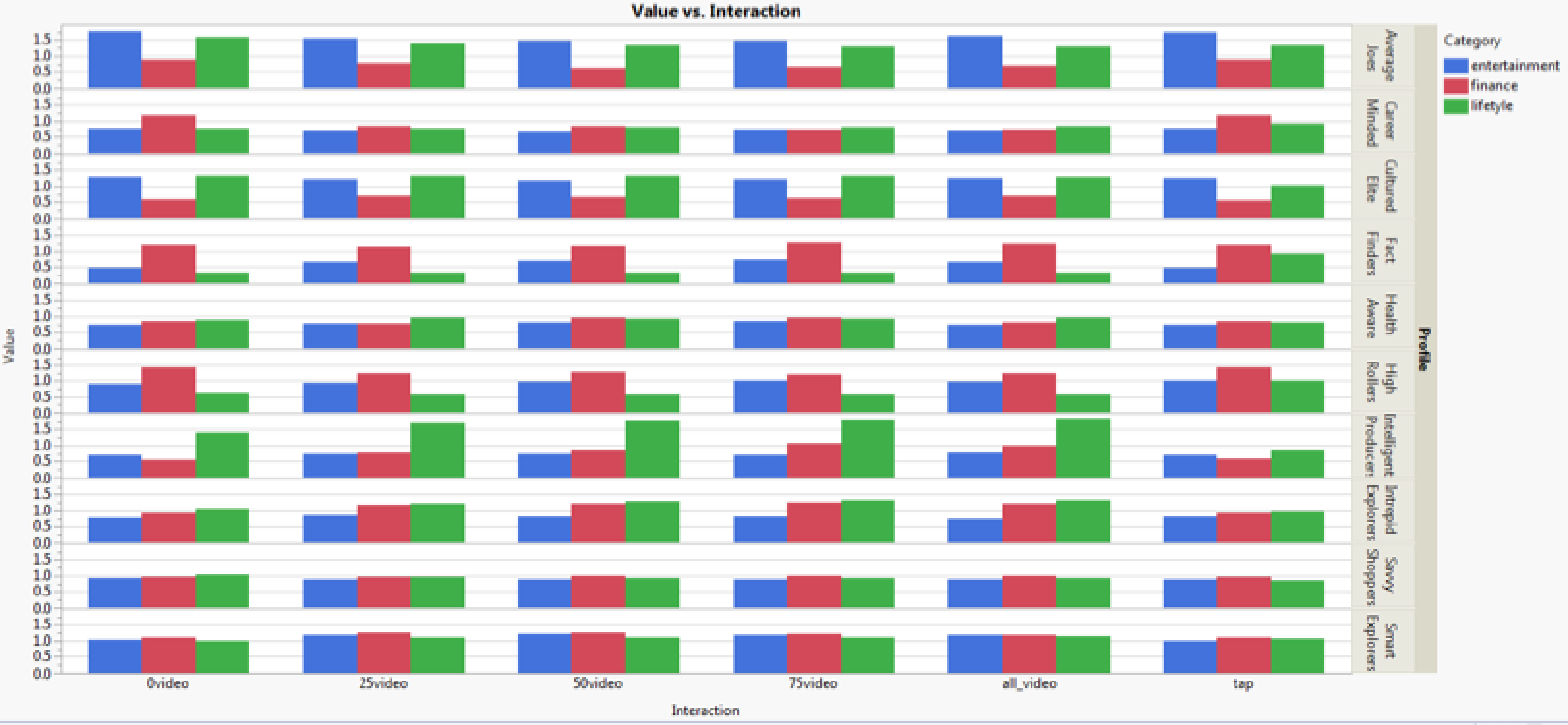}
\caption{The Index values for the interaction with ad videos across the ten profiles partitioned by ad category (finance-red, lifestyles-green and entertainment-blue). }
\label{int}
\end{figure*}

The k-means clustering identified ten clusters, see Figures \ref{fig:profiles}-\ref{fig:profiles2}.  The ten clusters and their corresponding centres are presented below.  As the clustering was performed on the z-score standardised data, a positive value indicates that the cluster profile have a higher than average percentage of their total apps installed consisting of that category and a negative value indicates that the cluster profile have a lower percentage of their total apps installed consisting of that category. The index values for the various video interactions across the ten different profiles when investigating the set of finance ads, the set of lifestyle ads or the set of entertainment ad are displayed in Figure \ref{int}.

The High Rollers profile, see Figure \ref{fig:high},  identified user with a greater than average percentage of apps in the iTunes categories games and sports and less than average percentage of apps in the social networking, lifestyles and finance categories.  These users tended to have numerous gambling apps suggesting many of these users are over 18.  The cluster centre for this profile suggests that users within the High Rollers profile are likely to use their mobile device for entertainment purposes rather than work.  The High Roller's users tended to have a high index value for finance video interactions (1.21-1.43) and a low index value for lifestyle video interactions (0.57 - 1.04), see Figure \ref{int}.  They had an average probability (index value 1.01) for tapping an entertainment ad.  When considering the profile this makes sense as gamblers are over 18, so they are more likely to be interested in finance.  In addition, due to their hobby they may have lots of money or be in financial trouble.  The index value for playing the finance video and getting to 25\% decreased from 1.43 to 1.23.  This may indicate that this profile is interested in finance ads but the particular ads investigated during this study were not suitable for many of the users.  

The Health Aware profile, see Figure \ref{fig:health},  contains users with a greater than average percentage of medical and weather apps making up their total apps.  These users also had slightly higher percentages of health/fitness, navigation, news and travel apps than the general population.  The percentage of their total apps corresponding to photo/video or social networking apps was lower than average.  This would indicate that these users are more health aware and tend to use their mobile device as a guide to maintaining or checking health and be active (navigation and weather apps suggest they may be outdoors often).  The Health Aware profile had index values of 0.73, 0.83 and 0.87 for tapping entertainment, lifestyle and finance ads respectively.  This suggests these users are generally  uninterested in all types of video adverts. However,  the index values for the completion of the video are 0.76, 0.95 and 0.81  for entertainment, lifestyle and finance ads respectively.  This shows that the Heath Aware profile users who tap a lifestyle advert are more likely than average to continue to watch it to completion.  This may suggest that the advert design may be putting off these users from tapping the ad and targeting these users by modifying the advert in a way to encourage them tapping a lifestyle ad may result in more of them watching the ad to completion.

The Fact Finders profile suggests these users are more likely to have reference apps than the general population,  see Figure \ref{fig:fact}.  This profile also had slightly higher percentages of their total apps in the categories entertainment, finance, games, lifestyles, sports and utilities.  This would suggest that these users tend to use their mobile device for both entertainment and finding facts.   They had lower than average percentages of their total apps from the categories music, photo/video and social media.  This could indicate that these users are likely to be active and around people as these apps would often be used when the user is alone. The Fact Finders had an index value of 1.23 for tapping a finance ad's video and this index value remained over 1 for the completion of the video, see Figure \ref{int}.  This suggests that this profile is interested in finance and should be targeted for future finance ads.  The Fact Finders had an index value of 0.92 for tapping a lifestyles ad but this dropped to 0.32 for playing the video.  This suggests these users are interested in lifestyle ads but the ads during this study were not suitable for them.  The index value for entertainment ad video interactions shows a unique dynamic.  The index value for tapping and playing an entertainment ad is low, 0.47 and 0.46 respectively, but this increased to 0.66 for playing the video to 25\%, 0.72 for getting to 50\% but reduced slightly to 0.65 for completing the video.  This indicates that the small number of checkers that click an entertainment ad's video are actually very interested and will watch the majority of the video, hence the index value increased as the video progressed.

The Career Minded, Figure \ref{fig:career}, are users that are interested in searching for jobs.  This was reflected by these users having a higher than average percentage of their total apps containing business apps, as the job search apps are considered business related.  These users may be unemployed or looking for a new challenge.  These users also had higher than average finance and reference type apps suggesting they use their mobile device for important aspects of their life. However, the lifestyle and food/drink was also higher than average, indicating they use their mobile device for lifestyle. This profile is interesting as it highlights how integrated into all aspects of life mobile devices have become to certain subpopulations. Their interaction with the various ad category videos is presented in Figure \ref{int}. The index values for tapping and playing a finance video were 1.20 and 1.18 respectively, however the index value for finance video progression decreased as the video progressed, resulting in a index value of 0.75 for video completion. This suggests the ad was promoting an unsuitable product for these users.  These users were less interested in entertainment and lifestyle ads with index values for interacting the the videos ranging between 0.76-0.91 for lifestyle and between 0.65-0.79 for entertainment.  As these users may be unemployed the result is as expected. They may be facing finance issues, which explains the interest in finance ads, and have less disposable money, which explains the lack of interest in lifestyle and entertainment ads.  

The Smart Explorers, Figure \ref{fig:smart}, may correspond to a group of users that often travel.  This profile cluster centre had a high value for navigation, utilities and news apps.  They also had slightly higher than average percentages of social media and entertainment apps when considering their total apps.  These users seem less interested in sports, lifestyle, finance, food/drink, productivity, music and reference.  The app profile may indicate that these users do not rely on their mobile device for everyday activities, with its main purpose as a phone but they may use it as a map at times.  The Smart Explorers had interaction index values between 1.01-1.14 for lifestyle, between 1.11-1.18 for finance and between 1.01-1.22 for entertainment.  The index value for finance video interactions was slightly higher than for lifestyles and entertainment but these users seem to be generally interested in all the ads.  This profile would be a good one to target for any type of advert.

The Average Joes profile corresponds to users that tend to use their mobile device for photo/video or social media only, see Figure \ref{fig:avg}.  These users had less than average percentages of apps out of their total apps from other categories. In particular, these users seem less likely to have entertainment, lifestyle or travel apps.  This profile may indicate these users are less tech savvy that on average.  The Average Joes had high index values for interacting with entertainment and lifestyle ad videos, with the tap index for entertainment being 1.75 and the tap index for lifestyle being 1.32.  They had low index values for interacting with finance ad videos with a tap index of 0.90 and the index as low as 0.69 for completing the whole finance video.  This shows the Average Joes have a preference for entertainment ads but also enjoy lifestyle ads.  They are slightly less likely than average to tap a finance ad, this indicates they did have some interest, however they are very unlikely to complete the whole finance video suggesting that the finance ads investigated were not aimed at them. Their complete interactions are displayed in Figure \ref{int}. 

The Intelligent Producers profile identified users who were more likely to have educational apps.  These users also had a higher percentages of their total apps corresponding to the iTines categories news, productivity and travel, see Figure \ref{fig:intelligent}.  This may indicate that these users tend to use their mobile device for education/learning activities and it is possible that these mobile devices are work ones.  The intelligent producers have an interaction dynamic that suggests these users are interested in lifestyle and know themselves, see Figure \ref{int}.  These users seemed less interested than average in entertainment ad videos with index values ranging between 0.69-0.78.  They are less likely than average to tap a finance ad's video, with an index value of 0.58. Interestingly, those that watch the video are likely to continue to completion with the index value for 25\%, 50\%, 75\% and 100\% showing a generally increases trend, with values of 0.78, 0.87, 1.08 and 1.00 respectively.  A similar trend was seen in lifestyles, where the tap index was 0.84 but the video completion was as high as 1.86. This shows that these users know themselves well and are able to judge whether an ad is suitable for them, so even though they are less likely than average to interact with a finance or lifestyle ad impression, those who chose to interact will engage with it until near the end of the video.

The Intrepid Explorers, see Figure \ref{fig:intrepid}, had a higher percentage of their total apps corresponding to productivity and travel apps.  These users also have higher percentages of news, books and entertainment.  This may indicate that these users tend to travel frequently and use their mobile device as a form of entertainment while travelling (reading books on it or other forms of entertainment).  This profile may correspond to young adults, as the travelling indicates that the users are adults and the use of book apps may indicate younger adults as older adults often dislike reading from small mobile devices due to their eye sight.  Interestingly, these users had lower percentage of social media, photo/video and games making up their total apps.  This may indicate these users enjoy socialising in person rather than though technology.  The Intrepid Explorers have a interaction dynamic similar to the Intelligent Producers but seem to be more interested in finance and less interested in lifestyles, see Figure \ref{int}.  These users had less than average tapping rate across all ad categories, with index values of 0.95, 0.91 and 0.80 for lifestyles, finance and entertainment respectively.  However, these users were more likely than average to continue watching the lifestyle and finance ad videos with index values for the video completion of 1.34 and 1.21 respectively.  The index values for entertainment ranged between 0.73-0.83 across all the video interactions. These users seem a good group to target any lifestyle and finance ads as they have high video completion index values.

The Savvy Shoppers, see Figure \ref{fig:savvy}, have a profile that suggests these users tend to use their mobile device for online shopping more than the general population.  These users tended to have a higher percentage of finance, food/drink, health/fitness and lifestyle apps making up their total apps than on average.  This suggests that the majority of this profile may be females who enjoy shopping and are body conscious. The Savvy Shoppers seem to have a low index value for interacting with all the ads.  Their index values ranged between 0.85-1.03 for lifestyles, 0.95-1.01 for finance and 0.88-0.91 for entertainment.  This indicates that the products being advertised where not appealing to these users in general or they are already aware of the product being advertised due to their interest in shopping.

The Cultured Elite profile, see Figure \ref{fig:cultured}, suggests these users use their mobile devices for listening to music.  The users in this profile generally have a higher percentage of their total apps corresponding to music, photo/video or social media apps.  This would indicate that these users have a preference for the arts.  These users are probably less likely that the average population to spend time on their phone for entertainment or lifestyle purposes such as shopping, watching videos or playing games. The Cultured Elite appear to be interested in lifestyles and entertainment ads but less interested in finance.  Their index values for interacting with finance ad videos ranged between 0.56-0.71, whereas their index value ranges for lifestyles and entertainment are 1.05-1.34 and 1.19-1.30 respectively.  These users could be targeted for lifestyle and entertainment ads but should not be targeted for finance ads in general. 

\begin{figure*}
        \centering
        \begin{subfigure}[b]{0.45\textwidth}
                \includegraphics[trim = 0mm 0mm 0mm 5mm, clip, width=\textwidth]{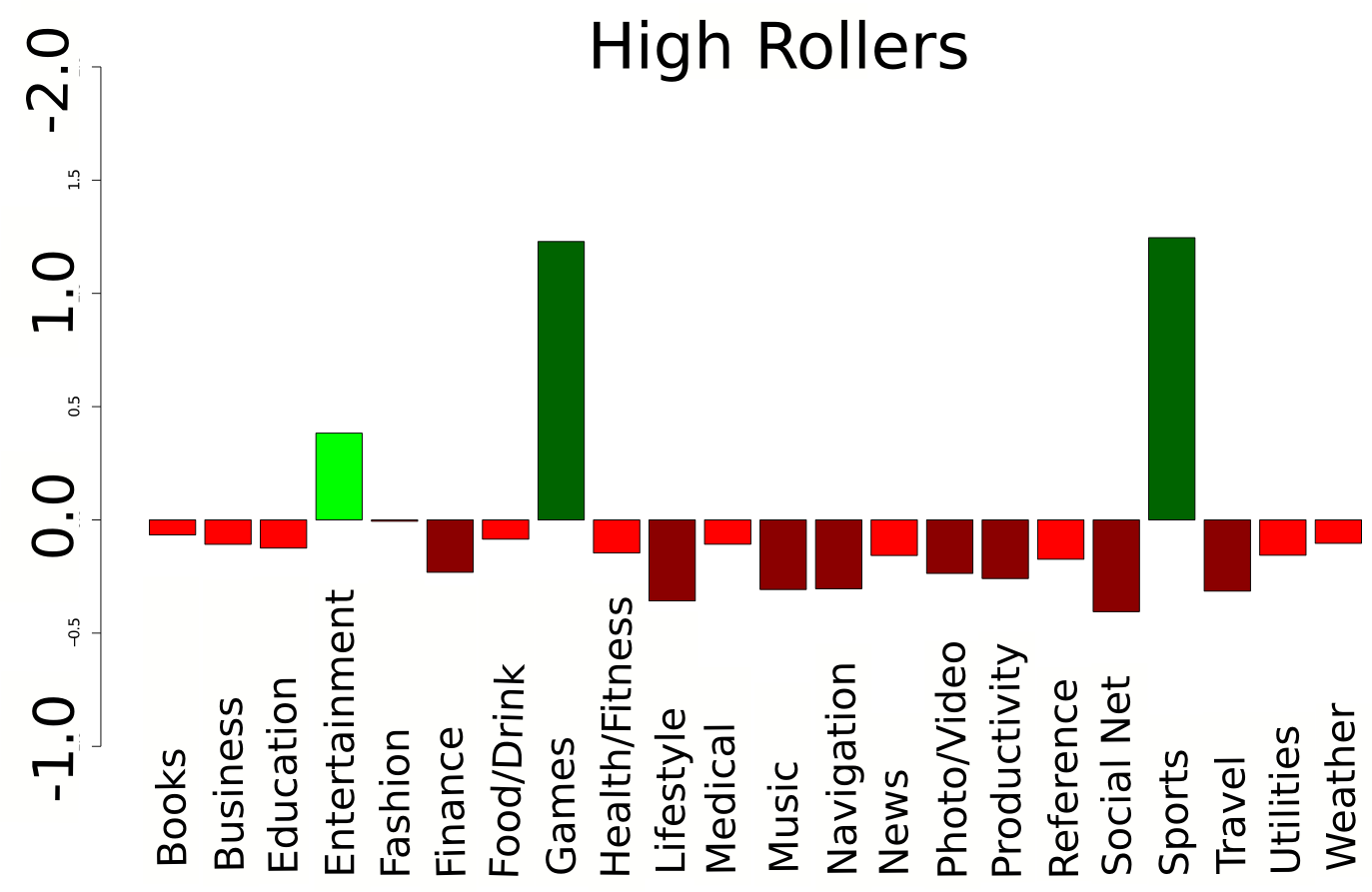}
                \caption{High Rollers}
                \label{fig:high}
        \end{subfigure}%
        ~ 
        \begin{subfigure}[b]{0.45\textwidth}
                \includegraphics[trim = 0mm 0mm 0mm 5mm, clip, width=\textwidth]{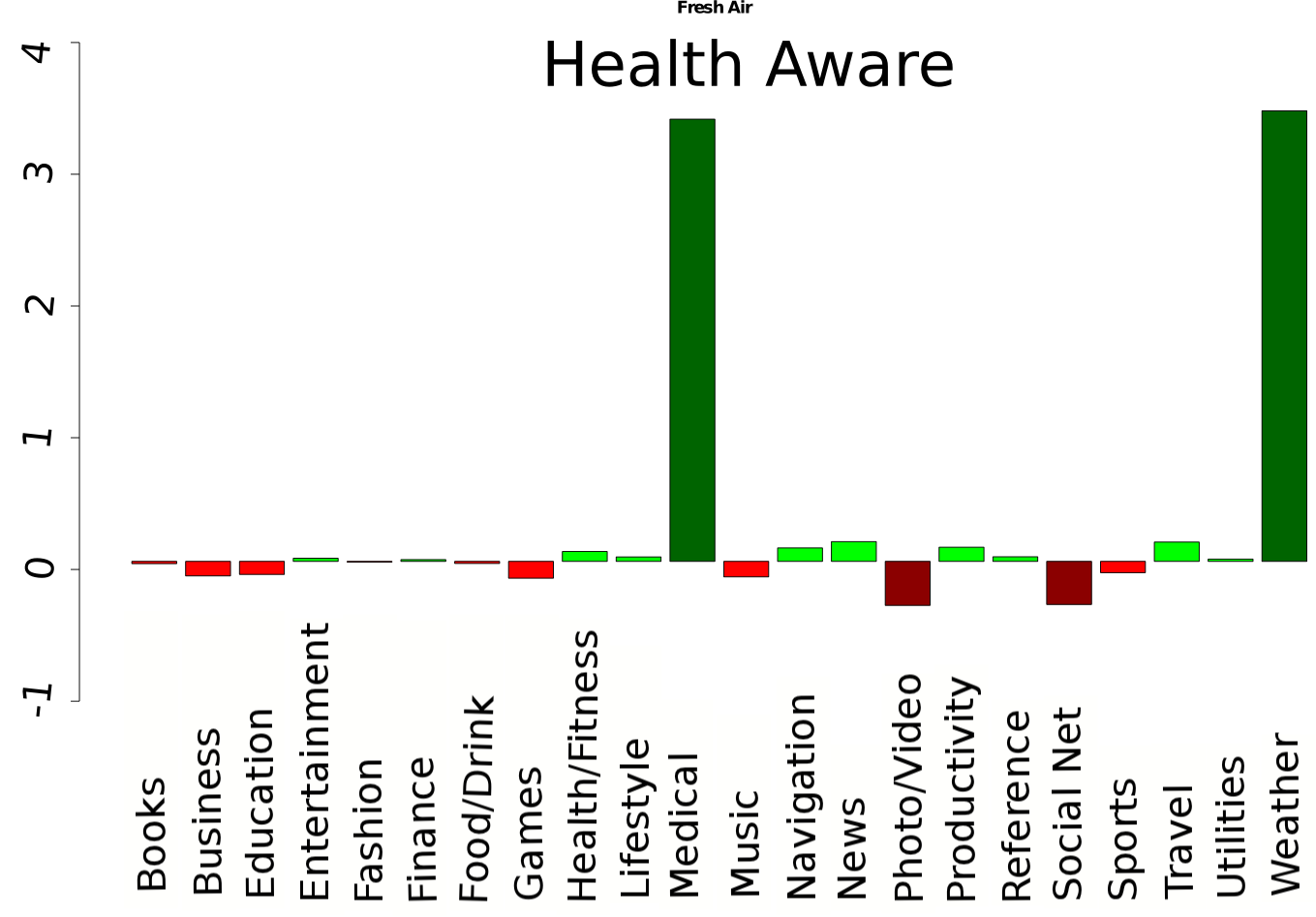}
                \caption{Health Aware}
                \label{fig:health}
        \end{subfigure}
        ~ 
        \begin{subfigure}[b]{0.45\textwidth}
                \includegraphics[trim = 0mm 0mm 0mm 25mm, clip,width=\textwidth]{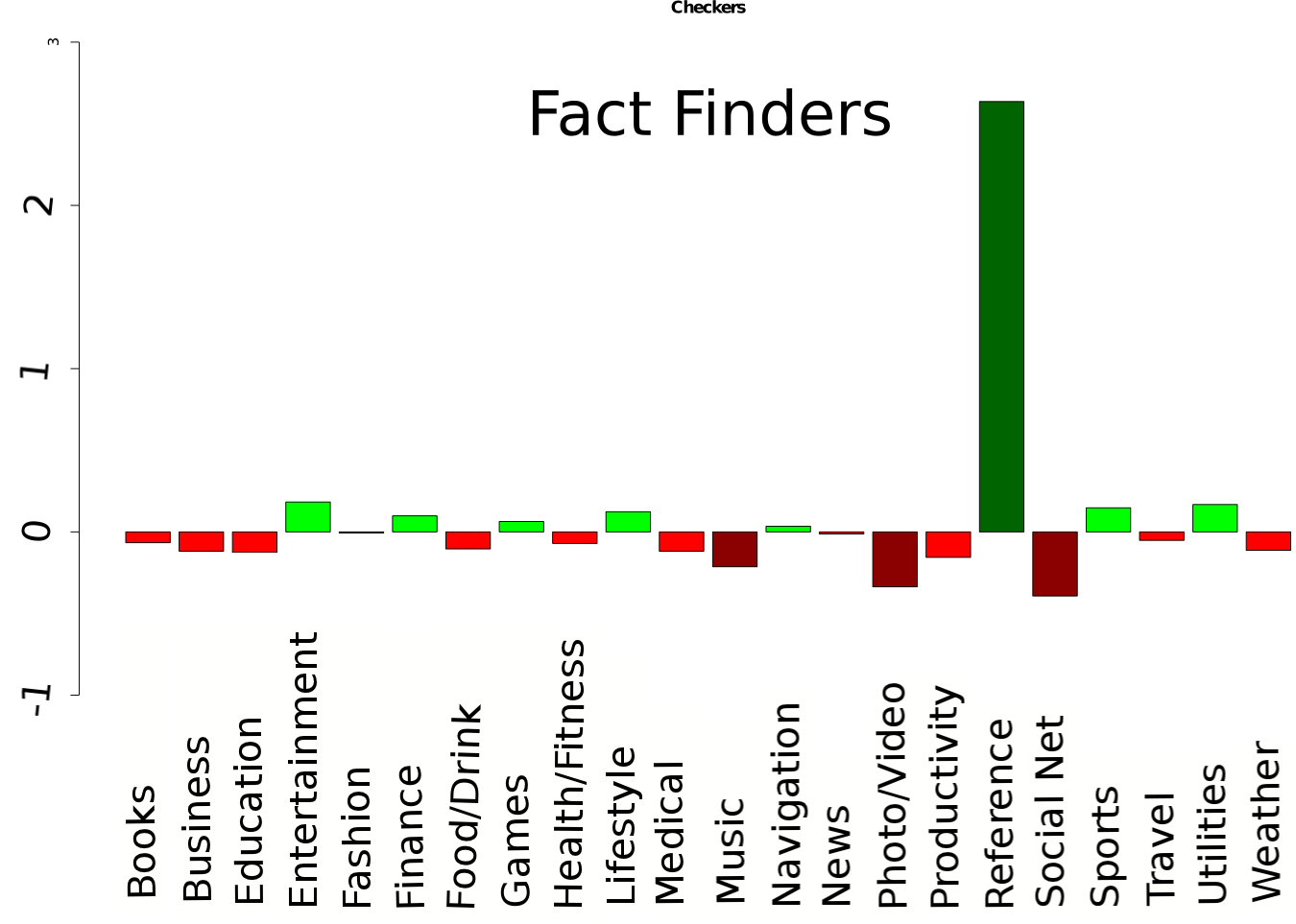}
                \caption{Fact Finders}
                \label{fig:fact}
        \end{subfigure}
        \begin{subfigure}[b]{0.45\textwidth}
                \includegraphics[trim = 0mm 0mm 0mm 15mm, clip,width=\textwidth]{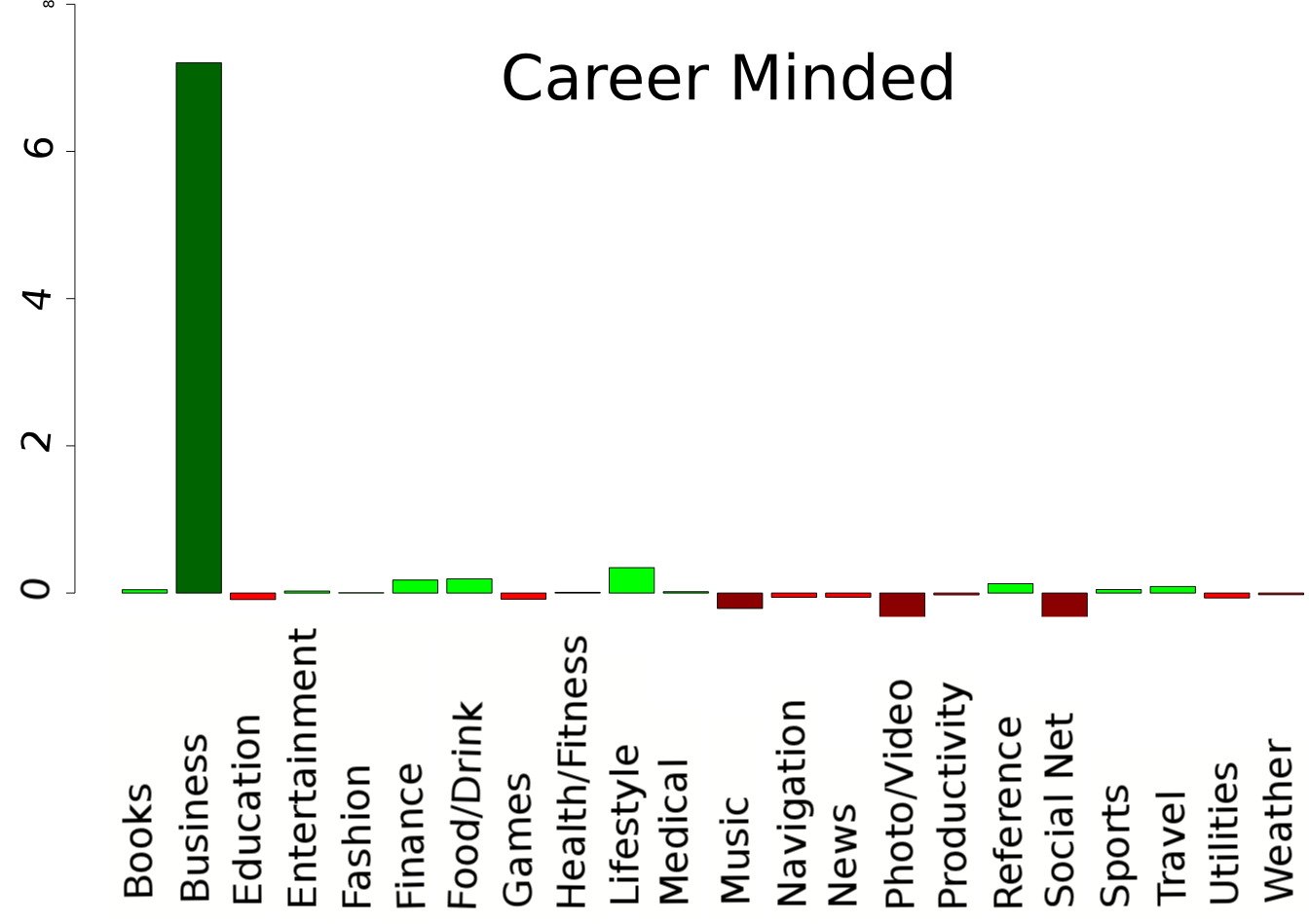}
                \caption{Career Minded}
                \label{fig:career}
        \end{subfigure}

        \begin{subfigure}[b]{0.45\textwidth}
                \includegraphics[trim = 0mm 0mm 0mm 5mm, clip, width=\textwidth]{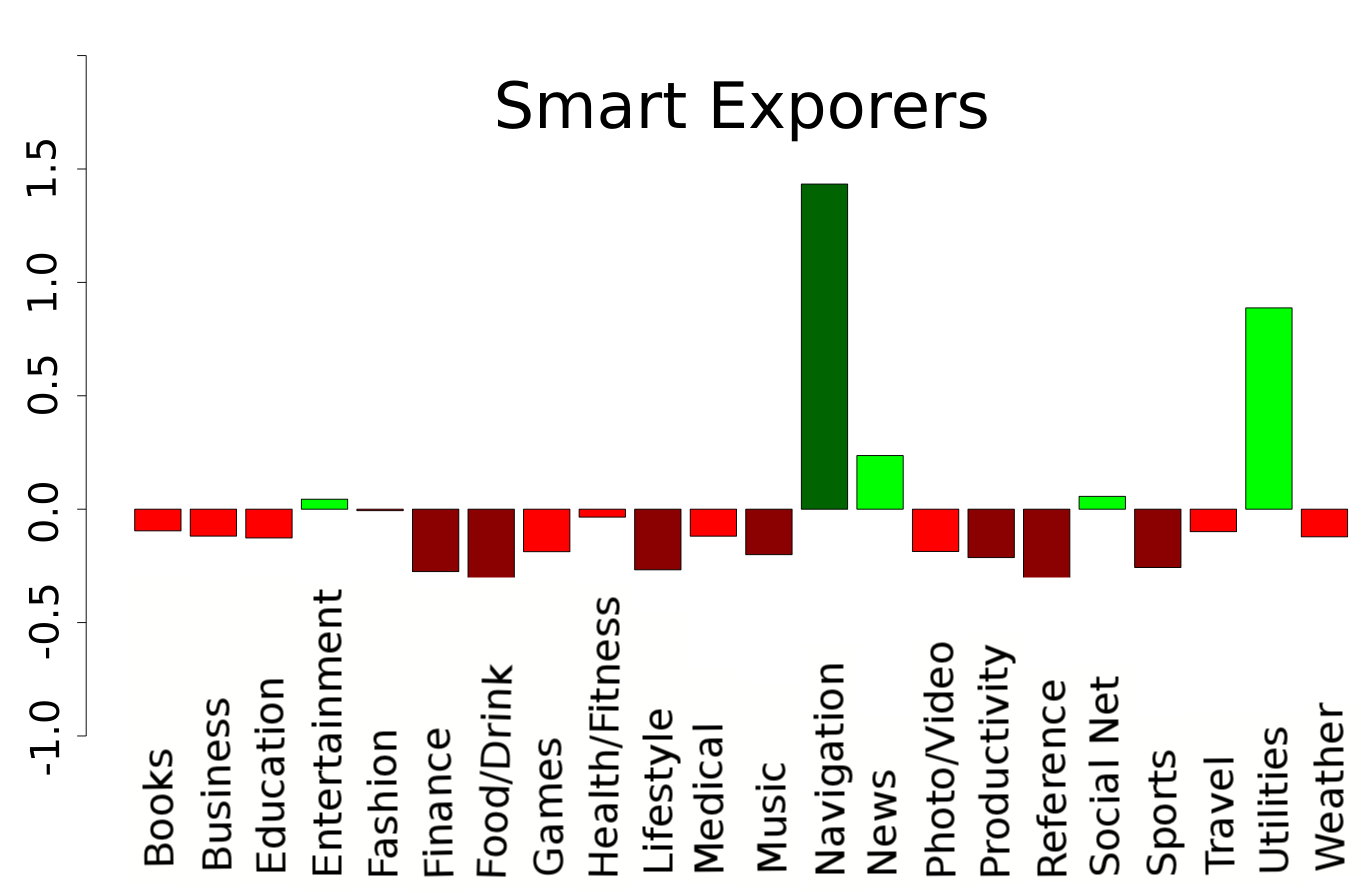}
                \caption{Smart Explorers}
                \label{fig:smart}
        \end{subfigure}
        ~ 
        \begin{subfigure}[b]{0.45\textwidth}
                \includegraphics[trim = 0mm 0mm 0mm 15mm, clip,width=\textwidth]{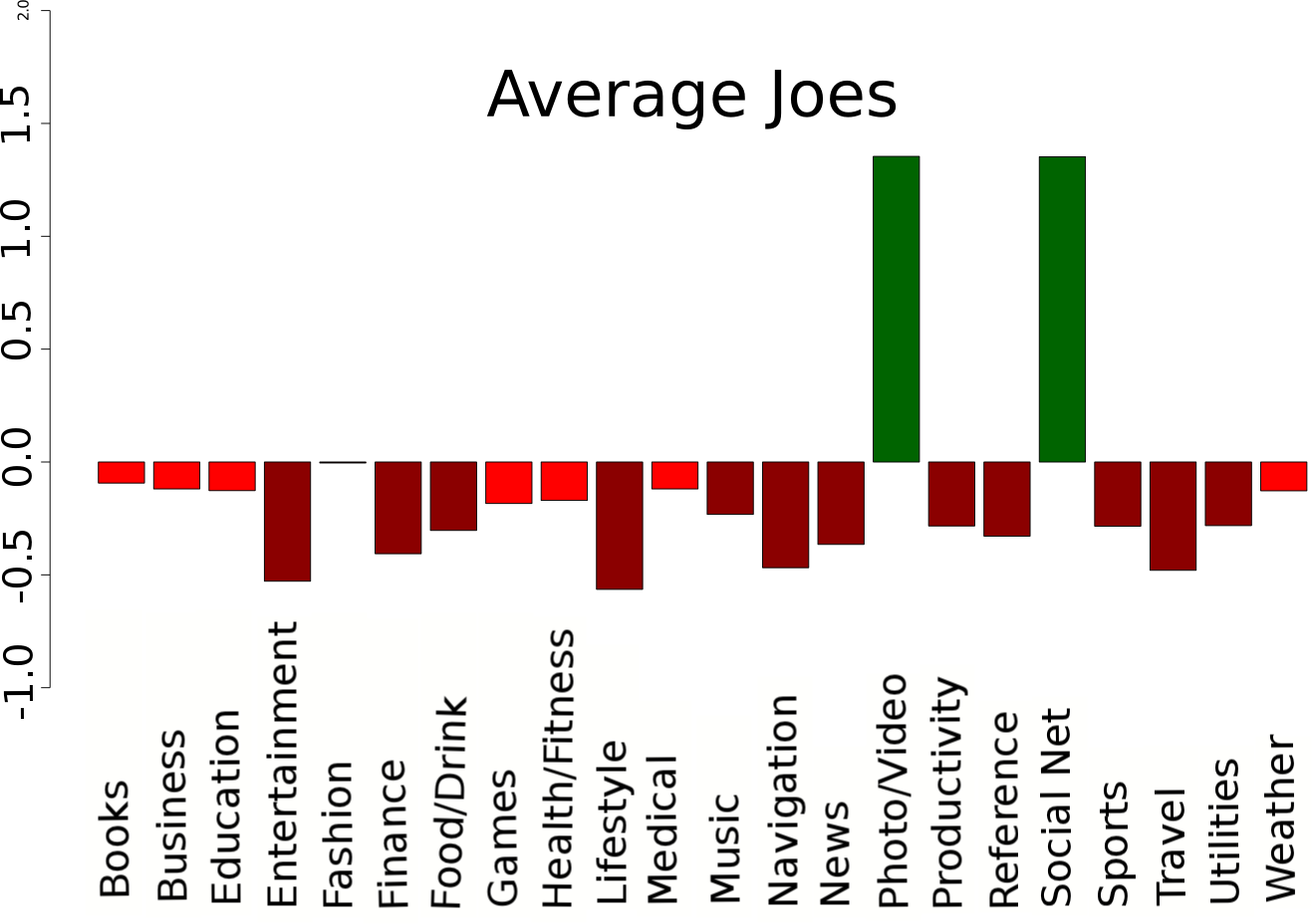}
                \caption{Average Joes}
                \label{fig:avg}
  \end{subfigure}
        \caption{Profiles of the different user clusters}\label{fig:profiles}
\end{figure*}

\begin{figure*}
        \centering
        \begin{subfigure}[b]{0.45\textwidth}
                \includegraphics[trim = 0mm 0mm 0mm 0mm, clip, width=\textwidth]{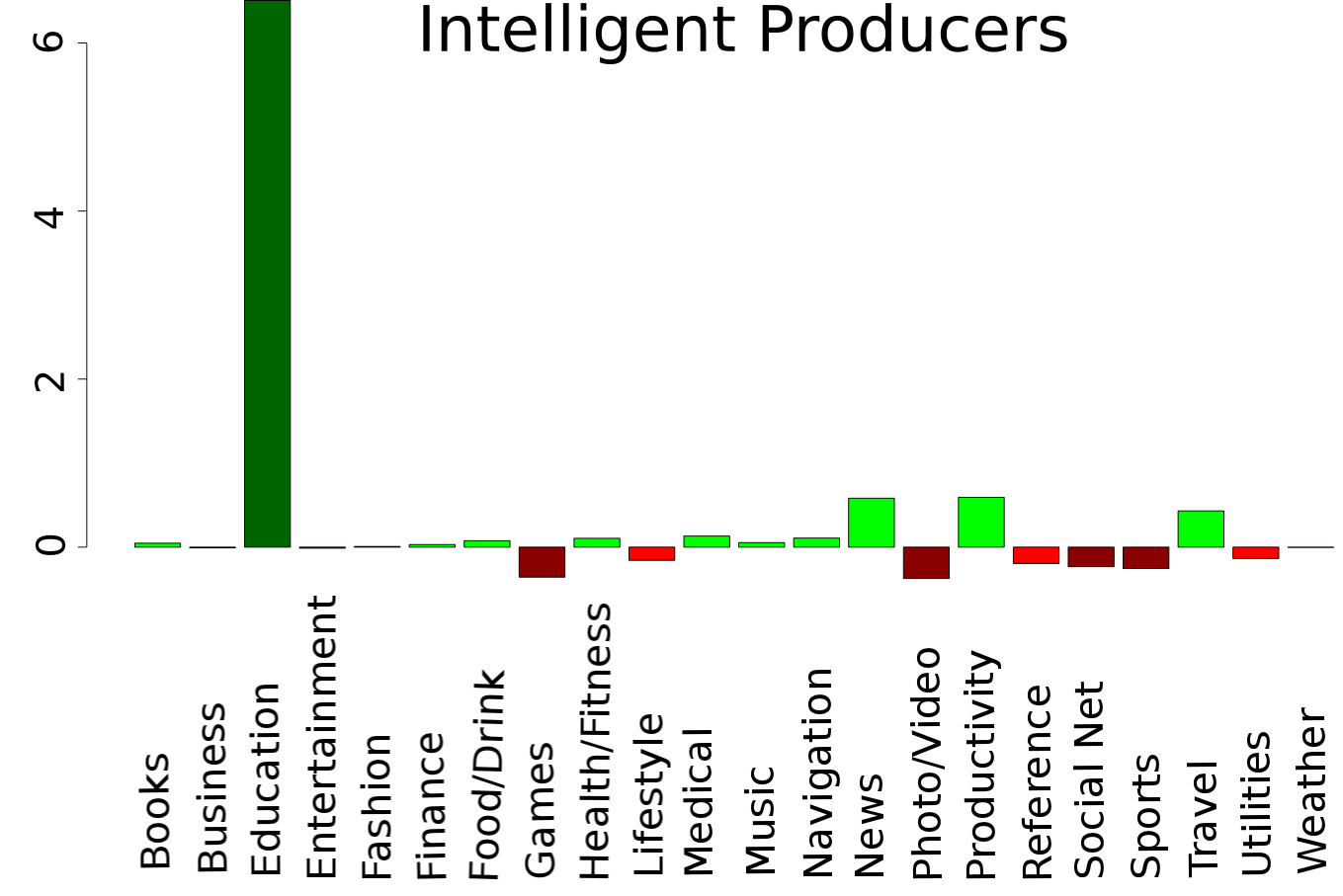}
                \caption{Intelligent Producers}
                \label{fig:intelligent}
        \end{subfigure}
        ~ 
        \begin{subfigure}[b]{0.45\textwidth}
                \includegraphics[trim = 0mm 0mm 0mm 0mm, clip, width=\textwidth]{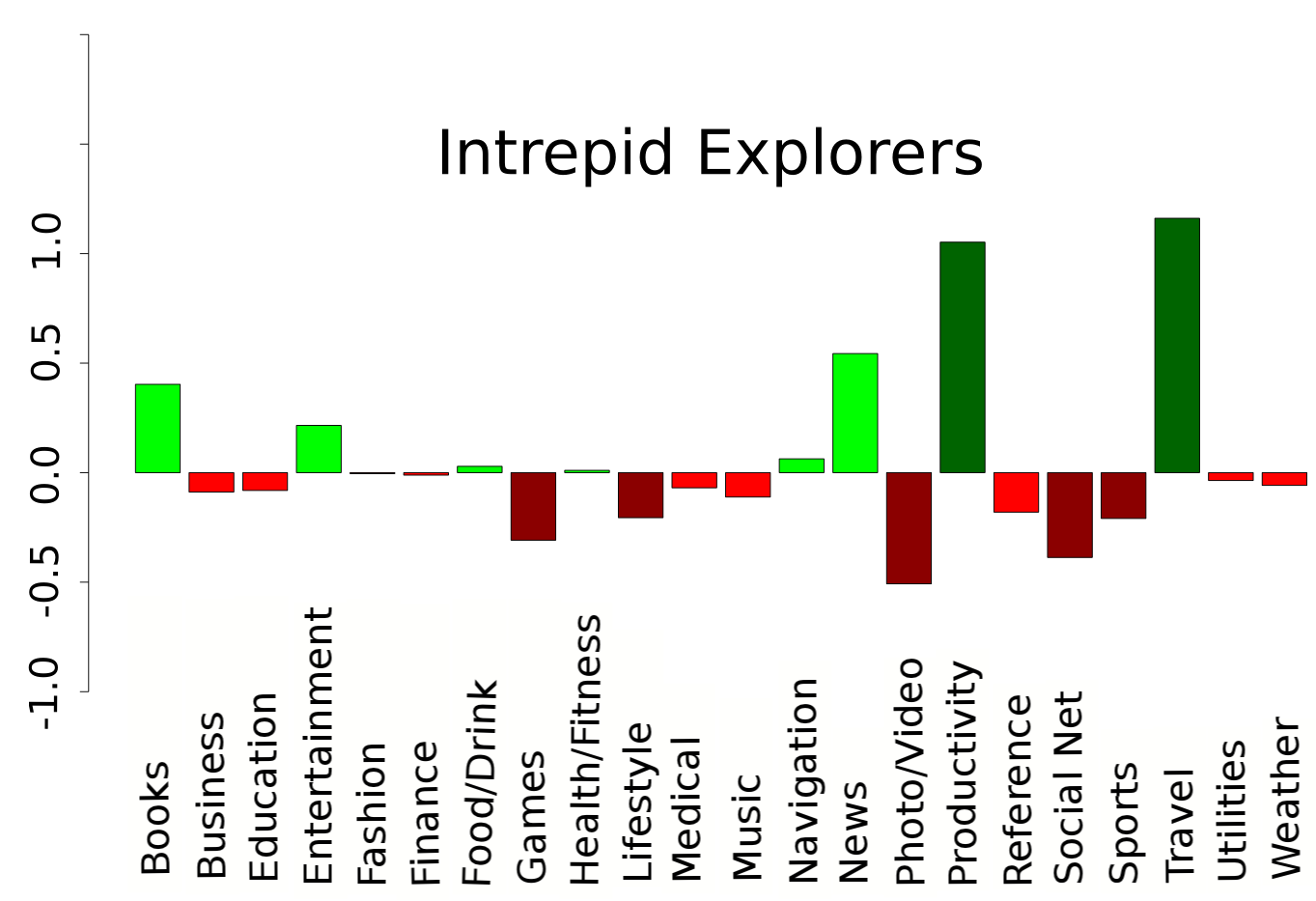}
                \caption{Intrepid Explorers}
                \label{fig:intrepid}
        \end{subfigure}
~ 
        \begin{subfigure}[b]{0.45\textwidth}
                \includegraphics[trim = 0mm 0mm 0mm 0mm, clip,width=\textwidth]{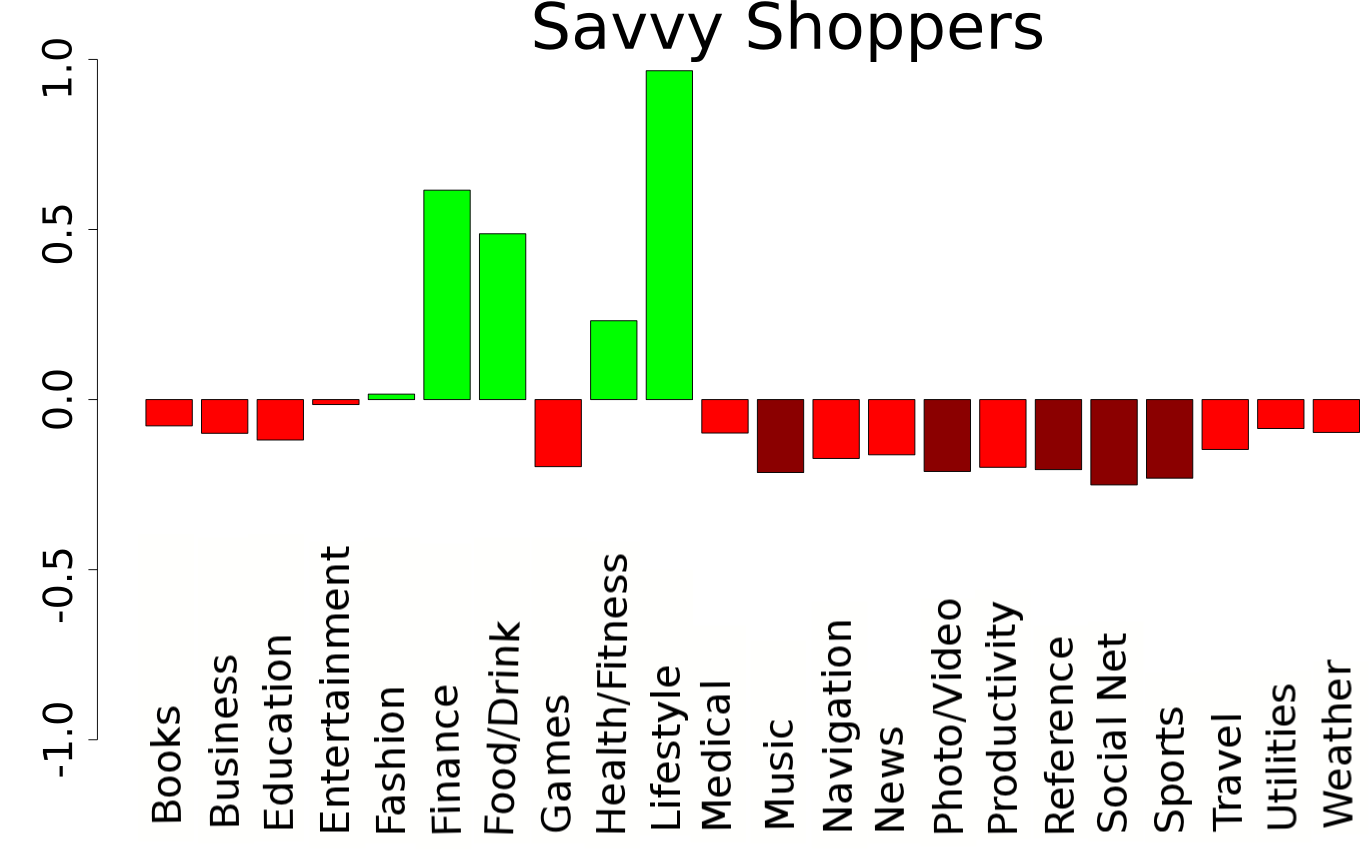}
                \caption{Savvy Shoppers}
                \label{fig:savvy}
        \end{subfigure}
~ 
        \begin{subfigure}[b]{0.45\textwidth}
                \includegraphics[trim = 0mm 0mm 0mm 0mm, clip,width=\textwidth]{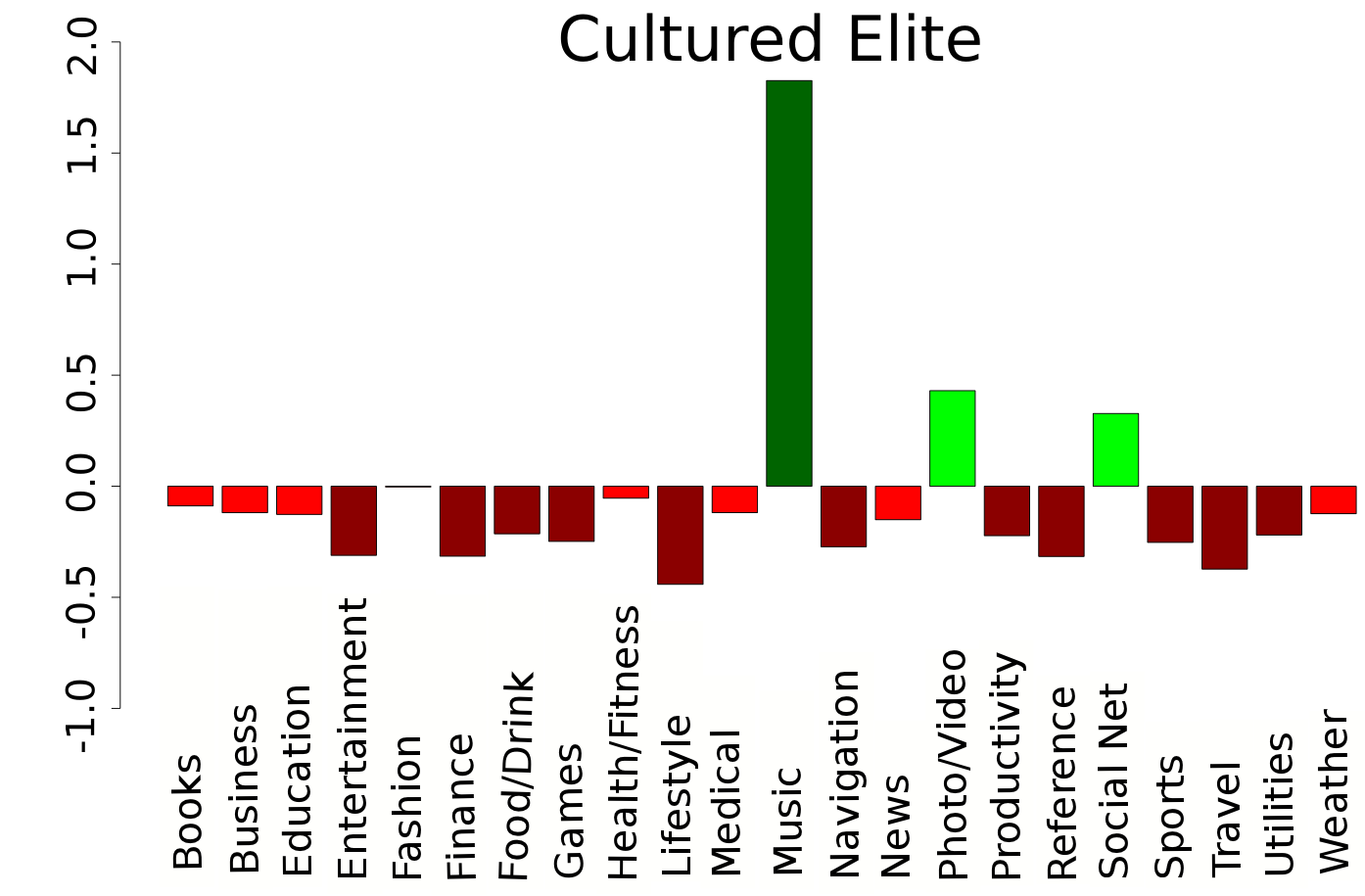}
                \caption{Cultured Elite}
                \label{fig:cultured}
        \end{subfigure}

        \caption{Profiles of the different user clusters}\label{fig:profiles2}
\end{figure*}

These results are interesting as it shows that the identified cluster profiles are a useful way to target ads but  simple targeting based on users having a high number of apps of the ad category may not always be suitable.  For example, the High Rollers had a low number of finance apps but they had a high index value for interacting with finance ad videos.  This result was also seen in the Intrepid Explorers, where they had a higher percentage of entertainment apps comprising their total apps than on average but had a index value less than 1 for entertainment ad video interactions.

Unsupervised learning is difficult to validate due to not knowing any true classes of users.  However, there are clear differences between the way the ten identified clusters interact with the three ad genres.  This provides external validation of the clustering performed during this study.  Therefore it seems suitable to use these clusters in future ad targeting to improve the impact of an advertisement campaign.

This is the first attempt to apply advanced computational methods for mobile advertisement targeting using a user's installed apps.  In this work we investigated ways to target a specific ad to users.  However, in future work,  if details about the ad were also recorded such as the size of the ad, the ad colour, the type (e.g., static, animated, video) and the target audience then these ad features could to be used to learn how to personalise an ad for a specific user.  For example, user A might be more likely to interact with an ad that is red than green or user B might prefer less intrusive ads.  By learning from historical impressions and interactions we could learn to adapt ads.  Rather than only serving ads to certain users, as promoted in this work, ads could be served to all the users but change style for the user.

\subsection{Association Rules}
The association rules containing the time of the day, the users profile and the number of apps a user has installed as an antecedent and a specified ad genre's interaction were identified. We mined 70 rules for finance ads where the consequence was play video, watch 50\% of video or watch 100\% of video. Examples of the rules are \{Class1, Career Minded\} $\to$ Play Finance Video with a left support of 1.4 $\times 10^{-4}$, confidence of 0.048 and lift of 2.8, \{Class1, High Rollers, daytime\} $\to$ Play 50\% of Finance Video with a left support of 0.011, confidence of 0.003 and lift of 3.1 and \{Class1, Health Aware, daytime\} $\to$ Play 50\% of Finance Video with a left support of 0.001, confidence of 0.004 and lift of 3.5. The lift value of a rule is similar to the index value, so a rule with a lift of 2 means that users with the rule's antecedent were two times as likely to do the consequence than in general. Therefore these rules tell us the time of the day to target a specific profile and whether to target the users within the profile with a high or low number of total apps.

We mined 55 rules for lifestyles ads where the consequence was play video, watch 50\% of video or watch 100\% of video. Examples of the rules are \{Class1, Intelligent Producers, daytime \} $\to$ Play Lifestyles Video with a left support of 1.6 $\times 10^{-4}$, confidence of 0.14 and lift of 2.3 and \{Class4, Average Joes, night\} $\to$ Play 50\% of Lifestyles Video with a left support of 1.4 $\times 10^{-5}$, confidence of 0.05 and lift of 2.1. 

For entertainment ads we mined 85 rules where the consequence was play video, watch 50\% of video or watch 100\% of video. Examples of the rules are \{Class1, Savvy Shoppers, daytime\} $\to$ Play 50\% of Entertainment Video with a left support of 0.004, confidence of 0.014 and lift of 2.5, \{Class1, Health Aware, night\} $\to$ Play 100\% of Entertainment Video with a left support of 1.1 $\times 10^{-4}$, confidence of 0.008 and lift of 2.7 and \{Class1, Average Joes\} $\to$ Play Entertainment Video with a left support of 0.030, confidence of 0.10 and lift of 2.

When personalising the mobile ad it is important to consider that there is a trade off between the number of users served the ad and the interaction rate. It would be simple to increase the interaction rate by personalised targeting but this may result in only a small percentage of users actually being deemed suitable for the ad. For example there are numerous rules identified where the lift is greater than 2, but the support of the rule is very small. If users were only given an ad impression based on these rules, then the frequency of impressions that resulted in an interaction would increase, but the number of users being given the impression will be very low. Therefore a set of rules need to be found that cover a sufficient number of users while also having an average lift greater than 1. This could be accomplished by using a metaheuristic algorithm in future work.

\section{Conclusions} \label{conc}
In this paper we describe an real life application of clustering and association rule mining that was used to analyse big data from the media industry.  The analysis found ten user profiles that could be utilised for personalised ad targeting to improve the performance of future advert campaigns.  Future could involve expanding the analysis to non-iPhone users, conducting a further study investigating how ads could be personalised for the user or investigating methods for finding a set of association rules that could be used to target users and improve performance while ensuring a sufficient number of users are served an impression.





\bibliographystyle{IEEEtran}
\bibliography{refs}
%

\end{document}